\documentclass[10pt,a4paper,twoside]{article}
\usepackage{epsfig}
\usepackage{baltlat6}
\usepackage{array}
\usepackage{here}
\usepackage{amssymb}

\begin{document}
\ \
\vspace{0.5mm}
\setcounter{page}{1}

\titlehead{Astronomy Letters, vol. \,00, XX--XX, 2015}

\titleb{FRACTAL PROPERTIES OF STELLAR SYSTEMS AND RANDOM FORCES}

\begin{authorl}
\authorb{O. Chumak}{1} and
\authorb{A. Rastorguev}{1,2}
\end{authorl}

\begin{addressl}
\addressb{1}{Sternberg State Astronomical Institute,  M. V. Lomonosov Moscow State\\
University,  Universitetskiy Prospekt 13, Moscow 119992, Russia;\\
chumako@sai.msu.ru, rastor@sai.msu.ru}
\addressb{2}{Physics Faculty, M. V. Lomonosov Moscow State University,\\
1 Bald. 2, Leninskie Gory, 119991 Moscow, Russia; rastor@sai.msu.ru}

\end{addressl}

\submitb{Received: 2015 October 26}
\begin{summary}The nearest neighbor distribution Chandrasekhar (1943) is generalized to fractal
stellar systems.For such systems an asymptotic distribution of the
magnitude of large random forces and a formula for the effective
mean interparticle spacing are derived. It is shown that in the case
of a power-law distribution of conditional density the derived
asymptotic fully agrees with the results obtained in terms of a
general approach. It is concluded that large random forces in a
fractal stellar medium are due entirely to the nearest neighbors
(clumps) located inside the sphere of the effective radius
determined from the generalized Holtsmark distribution.
\end{summary}

%% \resthead is the RUNNING TITLE at top of the pages
\resthead{Stellar systems and random forses}
{O. V. Chumak, A. S. Rastorguev}

\sectionb{1}{INTRODUCTION}
In this paper we investigate random forces acting on a test star
that are caused by multiscale stellar density fluctuations (fractal
stellar substructures). Currently, the fractal structure f
star-forming regions on spatial scales from 1 to 200-300 pc is well
established by observations of the young population in the Milky Way
and other galaxies (Efremov $\&$ Elmegreen, 1998; Elias et al.,
2009; Sanchez et al., 2010; Elmegreen et al., 2014; Gouliermis et
al., 2014). Our earlier analysis of the Geneva-Copenhagen survey of
FG-type dwarfs (Nordstrem et al., 2004; Holmberg et al., 2007, 2009)
revealed similar fluctuations with magnitudes substantially
exceeding those expected for uniform random (Poisson) space
distributions to be present in the solar neighborhood, i.e., in the
Galactic field. The fractal nature of the distribution of FG-type
dwarfs is believed to also reflect the features of the overall
distribution of local stars: we showed (Chumak $\&$ Rastorguev,
2015) that the space distribution of FG-type dwarfs in the solar
neighborhood differs significantly from uniform random distribution
on all available scale lengths (from 1 to 20 pc), i.e., that the
stellar medium demonstrates significant and  well-defined
"coarse-grained structure", which can be described by the average
Hausdorff dimension of $ D \approx 1.23$.
    One of the main problems in the study of the kinetics of a system of                                                                                                                                                                                      gravitating point masses is the analysis of the forces that act on test particles. In stellar dynamics such an analysis is traditionally performed by investigating the statistical and dynamic properties of the gravity force produced by the medium surrounding the test star. The results of such an analysis allow one to derive stochastic and kinetic equations of stellar dynamics in different approximations. The stellar dynamics eq                    uations derived in terms of such an approach may have different forms depending on the formulation of the problem and assessment of the effect of various factors. The fundamental nature of the problem of forces acting on a test star becomes clear from the fact that the corresponding equations form the basis for all the main results in stellar dynamics including the relaxation theory.
Analysis of random forces usually begins with selecting a model for
the medium surrounding the test particle. The simplest model is that
of infinite uniform static medium that is (1) isotropic on large
scales, (2) with limitedfluctuations, (3) consisting of point
particles of various masses randomly distributed according to
Poisson's law, and (4) without mutual correlations. Such a model is
known Chandrasekhar (1943) to imply the Holt
smark distribution of the magnitude of the random force acting on a
test star. Based on the Holtsmark distribution one can derive not
only the asymptotic formulas for the distribution of force, but also
compute the most probable random force for the model of the medium
considered. However, the second moment of the distribution
(dispersion) diverges, makings it difficult to construct consistent
kinetics of the stellar medium in terms of such a model. The
allowance for spatial correlations may in some cases result even in
the divergence of the first moment Gabrielli et al.(1999). As is
well known, the Fokker-Plank approximation also results in
divergences in the formulas for the force of dynamic friction and
components of the diffusion tensor, which can be prevented by
artificially constraining t
he Coulomb logarithm by imposing certain distance limits (the size
of the system or the average spacing between the stars). The classic
model mentioned above yields practically infinite relaxation times
exceeding the age of the system by several orders of magnitude.
These relaxation times are inconsistent with the results of
observations, which imply the well-mixed state of the stellar
population. Ogorodnikov (1958) called the problem of unrealistically
large relaxation times the fundamental paradox of stellar dynamics.
Thus this model, which yielded excellent results for quasi-neutral
uniform plasma, proved to be of little use in the case of
gravitating stellar medium where it can be considered only as a
first approximation despite the known similarity of the two types of
media. All this motivates the researchers to construct model
distributions of gravitating masses that would more adequately
describe the observed distribution. Such models are believed to
eventually eliminate the divergences and enable the development of
internally consistent kinetic theory of stellar medium that would
agree better with observational data. The class of anisotropic
models fits into this category. It is well known from observations
that stellar medium in stellar systems is almost always anisotropic
because of the presence of external field. Thus Chumak $\&$
Rastorguev (2014) showed that with the anisotropy of velocity
distribution taken into account the classic formula for relaxation
can be generalized resulting in strong reduction of the relaxation
time compared to standard theory. The second important factor that
determines the development of kinetic processes in stellar medium is
its structural inhomogeneity. Multiscaled correlation structures in
gravitating media have always been a particularly popular topic
among the researchers. The origin of these structures, their
physical and dynamic properties have been studied in such depth and
detail as permitted by the available observing techniques, physical
concepts, and the state of the development of the methods of
mathematical analysis. It is evident that young
 stellar groupings - open clusters and associations - may still
 contain residual structures that formed during the gravitational
 fragmentation of protoclusters from molecular clouds. These residual
 structures include binary and multiple systems, many of which will disrupt as a result of stellar encounters.
 However, because of the very nature of gravitational interactions between the stars, gravitationally unbound
 short-lived substructures may form in an initially quasi-uniform medium. The discovery of such structures in
 the solar neighborhood (Chumak $\&$ Rastorguev, 2015), i.e., among the rather old population of the Galactic disk,
 is indicative of the efficiency of this mechanism, which appears to be universal. If this is the case, estimating
 the effect of fractality on the kinetics of  stellar systems, i.e., its contribution to the distribution of
 random force, should be a relevant and interesting task.

\sectionb{2}{STOCHASTIC HIERARCHIES IN GRAVITATING MEDIA}

There is nothing unusual in the existence of fractal structures in the stellar medium. Carpenter (1938) found
the number of galaxies in a cluster, $N(r)$, to depend on $r$ as $\sim r^{1.5}$ rather than $r^3$, as would be
expected for uniform space distribution of galaxies. In other words, the number density of galaxies varies as
$n(r) \sim N(r)/r^3 \sim r^{-1.5}$, i.e., it decreases with increasing characteristic size in accordance with
a fractional-power law with exponent (-3/2). Carpenter concluded that clusters of galaxies are not quasi-isolated
objects, but rather sort of concentrations in some non-uniform but non-random limited-density distribution.
De Vaucouleurs (1970) repeated Carpenter's analysis using more up-to-date data (de Vaucouleurs et al., 1991) and
concluded that the distribution of galaxies, on the average, obeys the law $n(r) \sim r^{-1.7}$; he generalized
Carpenter's conclusions extending his results not only to clusters of galaxies, but also to the entire galaxy medium.
After de Vaucouleurs (1970) paper and many other studies published since then by different authors the fractal
structure of the world of galaxies became the generally accepted model in extragalactic astronomy (see, e.g., Wu et al., 1999).
It turned out that the world of galaxies is organized hierarchically, and therefore any observer associated with
an object that is a part of this hierarchy sees the average density around him to decrease with distance.
At the same time, there is no special location in this medium, i.e., any sufficiently large and equal volumes
have the same average density irrespective of their relative location. This density can be referred to as invariant
conditional density (CD). In the case of synchronous variation the size of these CD volumes varies with characteristic size $r$ as
$$
n(r)\sim r^{-\alpha}   \eqno(1)
$$
where $\alpha \approx 1.7$. This is what is called the Carpenter-de Vaucouleurs law of galaxy clustering.
Based on the results of de Vaucouleurs, Mandelbrot (1977, 1988) interpreted the law represented by eq. (1)
as a special case of stochastic self-similarity of three-dimensional fractal sets, which obeys relation:
$$
N(r)\sim r^{3-\alpha} = r^{D}   \eqno(2)
$$                   .
Here $D$ is referred to as the fractal (Hausdorff) dimension of the medium. Mandelbrot showed that depending
on the properties of the medium such fractal structures can occur in different gravitating media and may have
 fractal dimensions in the $0\leq D \leq 3$ interval, i.e., exponent $\alpha$ may vary in the interval from -3 to 0.
Numerous numerical simulations of the dynamic evolution of
self-gravitating systems performed using Nbody algorithms also
confirmed the formation of stochastic hierarchical structures in
initially uniform self-gravitating systems. These hierarchical
structures are similar to those actually observed both in the galaxy
and interstellar media over a range of spatial scales spanning
several orders of magnitude (see, e.g., Semelin aand Combes, 2000,
2002; Bottaccio et al., 2003, and others). In our earlier study
(Chumak and Rastorguev, 2015) based on an analysis of the space
distribution of stars in  the Geneva-Copenhagen survey, we found
that the solar neighborhood in the Galaxy has a fractal structure
with a mean Hausdorff dimension of $D \approx 1.23$. We also
estimated the dependence of conditional density on distance from
each star of the sample, namely, we determined the number $N(r, i)$
of stars inside spheres of increasing radii $r$ centered on i-th
star of the sample, and then computed the corresponding conditional
density $n(r, i)$. It is evident that for each radius r there are
some maximum, mean, and minimum values of conditional density. For
the solar neighborhood these three series of $n(r)$ values can be
with high degree of confidence fitted by power laws:
$$
n(r) = hr^{-\alpha}   \eqno(3)
$$
The parameters of dependence (3) are equal to $h \approx 0.036$
and $\alpha\approx 0.948$ (at a Pearson confidence level of $R^2 \approx 0.867$)
for the series of maximum conditional density values $n_{max}(r)$; $h \approx 1.644$
and $\alpha \approx 1.769$  ($R^2 \approx 0.991$) for the mean conditional density $n_{mean}(r)$,
and $h \approx 0.483$ and $\alpha \approx 3.007$ ($R^2 \approx 0.9998$) for the minimum conditional density $n_{min}(r)$.

The existence of minimum conditional density values indicates that in the stellar medium
there always regions where the mass inside a sphere surrounding a star is practically
independent of the radius of this sphere ($\alpha \approx 3.0$). In other words, voids
occur on all the scales considered (from 1 to 20 pc), indicating the presence of intermittency
in the stellar medium.

    The average Hausdorff dimension ($D = 3-\alpha$) of the space distribution of the
FG-type dwarfs studied is equal to $D \approx 1.23$. Figure 1 shows the log-log plot
of the distance dependence of conditional density averaged over all stars of the sample
and corresponding to this Hausdorff dimension.
\begin{figure}[!tH]
\vbox{\centerline{\psfig{figure=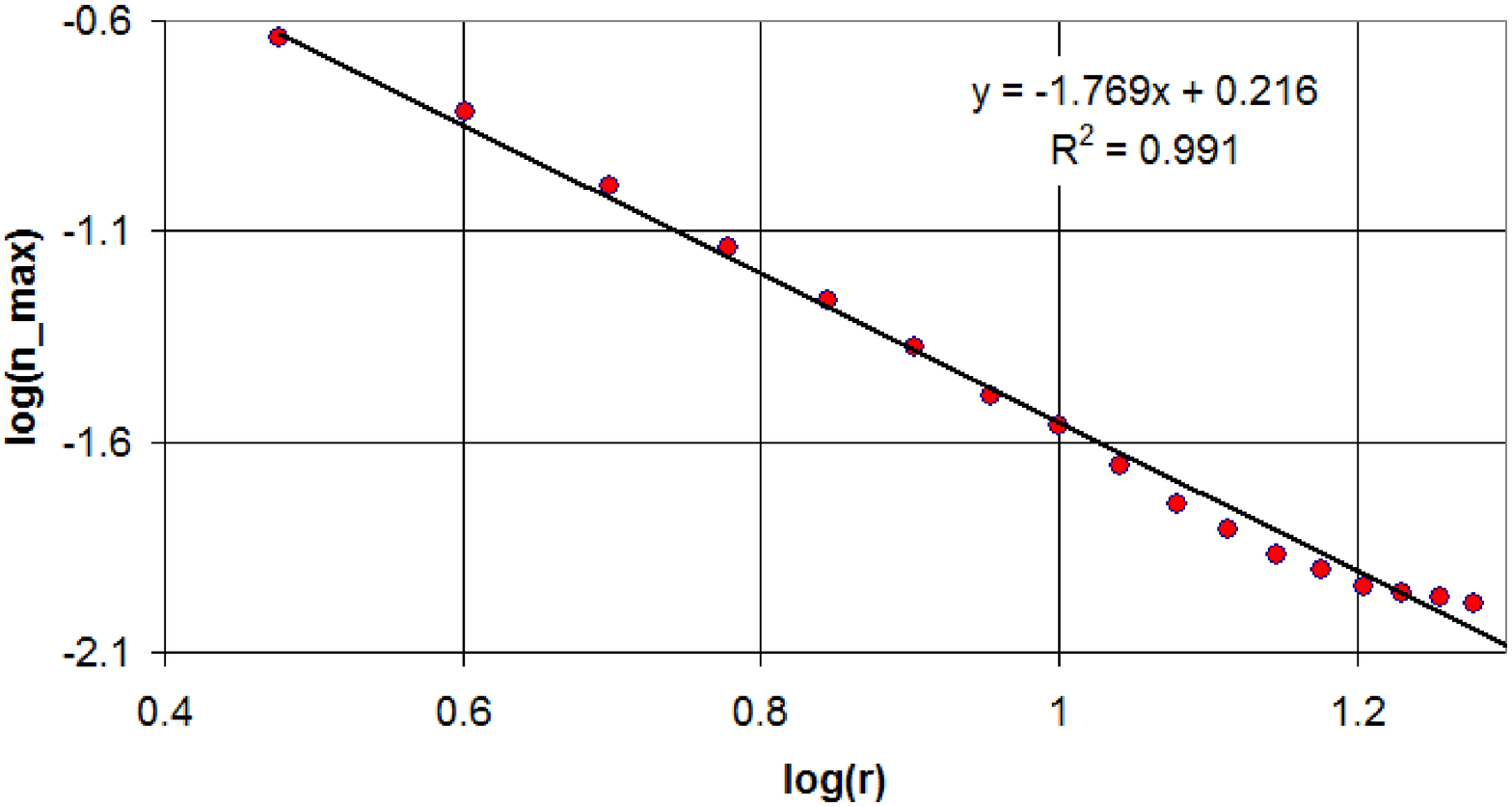,width=100mm,angle=0,trim=30mm
0mm 30mm 0mm }} \vspace{1mm} \captionb{1} {Average dependence of the
logarithm of conditional density on the logarithm of the radius of
the volume based on the data of the Geneva-Copenhagen survey} }
\end{figure}

Given the lack of similar data for the space distribution of other stars in the local
volume we assess the effect of fractality on stellar kinetics assuming that the fractal
properties of the local sample of FG-type dwarfs are representative of the entire stellar
population of the Galactic disk.

\sectionb{3}{ASYMPTOTIC LAW OF THE DISTRIBUTION OF THE MAGNITUDE OF RANDOM FORCE IN A
FRACTAL MEDIUM IN THE NEAREST-NEIGHBOR APPROXIMATION}

As we pointed out above, Chandrasekhar (1943) derived an exact
solution for the distribution of the magnitude of fluctuating random
force in the case of uncorrelated (Poissonian) static random space
distribution of stars. This solution for the magnitude of the force
has the form of the Holtsmark distribution:
$$
W(|\overrightarrow{F}|) = H(\beta)a^{3/2}   \eqno(4)
$$
where
$$
H(\beta)=\frac{2}{\pi\beta}\int_{0}^{\infty}exp[-(x/\beta)^{3/2}]x\sin(x)dx
\eqno(5)
$$
and in the case of equal-mass stars
$$
a =(4/15)(2\pi G m)^{3/2} n            \eqno(6)
$$
The dimensionless force is equal to
$$
\beta =|\overrightarrow{F}|/a^{2/3}           \eqno(7)
$$
Chandrasekhar (1947) showed that the asymptotics of the Holtsmark
distribution in the approximation of large random forces exactly
coincides with the distribution of the random force acting on a
unit-mass test star and due to the nearest neighbor of mass $m$
located at distance $r$:
$$
|\overrightarrow{F}|=\frac{G m}{r^2}          \eqno(8)
$$
The asymptotic formula for $W(|\overrightarrow{F}|)$ can be derived
from the nearest neighbor distance distribution $w(r)$:
$$
w(r)dr = 4\pi n exp(-4\pi r^3 n/3)r^2dr          \eqno(9)
$$
$$
W(|\overrightarrow{F}|)d|\overrightarrow{F}| = 4\pi n(Gm)^{3/2}
 exp[4/3 \pi
n(Gm)^{3/2}|\overrightarrow{F}|^{-3/2}]|\overrightarrow{F}|^{-5/2}
d|\overrightarrow{F}|   \eqno(10)
$$
We now generalize the Holtsmark law to the case of fractal
distribution assuming that the force acting onto the test star is
also determined by its nearest neighbor exclusively and taking into
account the clumpy structure of the stellar medium. To this end, the
distribution of the nearest-neighbor distance should be derived
using the distance dependence of average density in the form of
formula (3):
$$
w(r) = [1 - \int_{0}^{r} w(r)dr]4\pi r^2 n(r) = 4\pi h [1
-\int_{0}^{r} w(r)dr]r^{D-1}           \eqno(11)
$$
implying that:
$$
\frac{d}{dr}[\frac{w(r)}{4\pi h r^{D - 1}}]= -4\pi h  r^{D -
1}\frac{w(r)}{4\pi h r^{D - 1}}           \eqno(12)
$$
According to formula (11),$w(r)\rightarrow 4\pi h r^{D-1} $  at $r
\rightarrow 0$ , and it therefore follows from equation (12) that
$$
w(r)dr = 4\pi h exp(-\frac{4\pi h}{3}r^D)r^{D-1}dr \eqno(13)
$$
This is the analog of formula (10), i.e., the generalization of the
nearest-neighbor distribution for the case of fractal mass
distribution. Further, given formula (8), we derive from
equation (13), after simple rearrangement, the following
generalization of formula (10) for the case of power-law density
distribution (3):
$$
W(|\overrightarrow{F}||D)d|\overrightarrow{F}| = 4\pi h
(Gm)^{D/2}exp[-\frac{4\pi
h}{3}(\frac{Gm}{|\overrightarrow{F}|})^{D/2}]|\overrightarrow{F}|^{-\frac{D+2}{2}}d|\overrightarrow{F}|\eqno(14)
$$
where, as above, $D$ is the fractal dimension.

In the limit of uniform medium $(\alpha \rightarrow 0, D \rightarrow
3, h \rightarrow n)$ formula (14) goes over into the standard
Holtsmark distribution (10).

In the limit of strong fields $( |\overrightarrow{F}|
\rightarrow\infty)$ we obtain from formula (14) the following
formula for the magnitude of random force $F =
|\overrightarrow{F}|$:
$$
W(F|D)\approx  4\pi h (Gm)^{D/2} F^{-\frac{D+2}{2}} \eqno(15)
$$

We now denote the dimensionless argument of the exponential function
in formula (14) as $x =\frac{4\pi h}{3}(\frac{Gm}{F})^{D/2} $ to
rewrite formula (14) in the form
$$
W(F|D) F = 3 x e^{-x}         \eqno(16a)
$$

or

$$
A W(F|D) = (3 x)^{(D+2)/D} e^{-x}         \eqno(16b)
$$

where
$$
A = \frac{(4\pi h)^{2/D}}{Gm}        \eqno(16c)
$$

Figure 2 shows the standard Holtsmark law (the lower curve) and our
derived distribution (the upper curve) that takes into account the
fractal nature of the distribution of local stars with the Hausdorff
dimension $D \approx 1.23$ (the upper curve).

\begin{figure}[!tH]
\vbox{\centerline{\psfig{figure=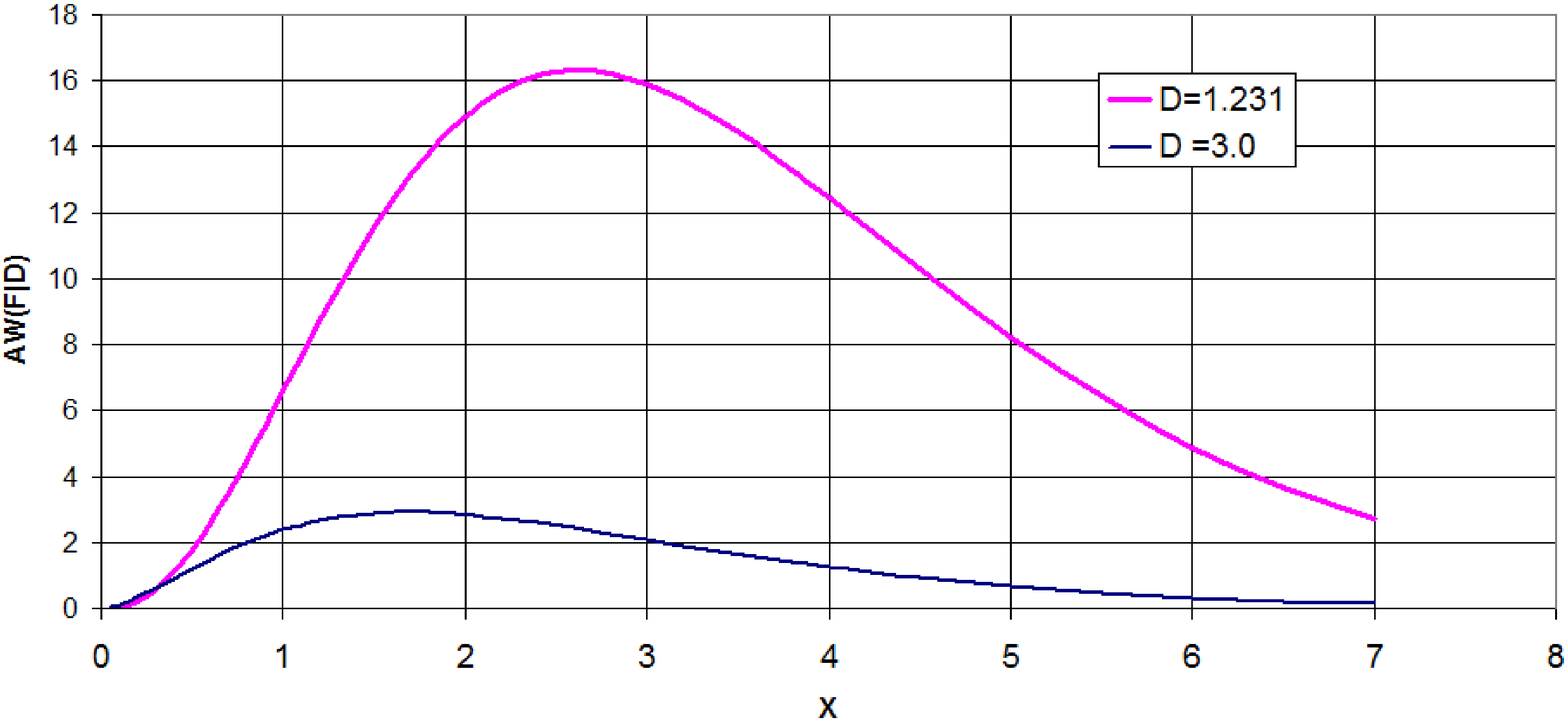,width=100mm,angle=0,trim=30mm
0mm 30mm 0mm }} \vspace{1mm} \captionb{2} {Average dependence of the
logarithm of conditional density on the logarithm of the radius of
the volume based on the data of the Geneva-Copenhagen survey} }
\end{figure}

It is evident from formulas (16) that probability distribution
$W(F)$ decreases most rapidly with force magnitude in the case of
uniform distribution ($D = 3$). In this sense the Holtsmark random
distribution is the lower limit for $W(F)$. Any real distributions
of gravitating masses with the mean conditional density of the form
of equation (3) with $\alpha > 0$  ($D < 3$) should generate more
slowly decreasing random force distributions $W(F)$, thereby
demonstrating that strong fields play more important role in the
kinetics of gravitating medium. Furthermore, it is evident from Fig.
2 that the most probable normalized random force is substantially
(by a factor of about 1.7) greater than the value computed for the
standard Holtsmark distribution.

The distribution of random force acting on a test star in the solar
region can be computed in the nearest neighbor approximation by
substituting into formula (14) the Hausdorff dimension $D \approx
1.23$ determined above. To apply distribution law (14) to other
parts of the Galaxy and to open and globular clusters, we should
ensure that power law (3) holds for these regions. Needless to say
that fractal dimension $D$ differs in each particular case.

\sectionb{4}{ESTIMATING THE EFFECTIVE INTERPARTICLE SPACING FOR THE FRACTAL MEDIUM}
The {\it effective} interparticle spacing rm and characteristic
effective time tm of mutual pairwise encounters in a fractal
gravitating medium can be determined from formula (13) for the
distribution of the nearest neighbor distance. By definition
$$
r_m = \int_{0}^{\infty}r w(r) dr   \eqno(17)
$$
where probability $w(r)$ should be computed by formula (13). Note
that for convenience of further computations $w(r)dr$ is better to
rewrite in the form
$$
w(x) dx = \frac{3}{D}e^{-x}dx       \eqno(18)
$$
where $x = \frac{4\pi h}{3}r^D$, $D = (3 - \alpha)$ is the Hausdorff
fractal dimension, and $h$ is the density parameter in formula (3).
Given formula (18), we then derive from formula (17) the following
exact estimate for $r_m$
$$
r_m = \frac{3}{D}(\frac{4\pi
h}{3})^{-1/D}\int_{0}^{\infty}e^{-x}x^{1/D}dx =
\frac{3}{D}(\frac{3}{4\pi h})^{1/D}\Gamma(\frac{D+1}{D}). \eqno(19)
$$

In the limit $D \approx 3$, $h \approx n$ - (uniform random Poisson
medium) we obtain the well-known Chandrasekhar's (1943) estimate for
the mean interparticle spacing
$$
r_m = (\frac{3}{4\pi n})^{1/3}\Gamma(\frac{4}{3})\approx 0.554
n^{-1/3}.           \eqno(20)
$$

Substituting our parameter estimates for the space distribution of
the near-solar sample of FG-type dwarfs, $h \approx 1.644$ and $D
\approx 1.23$, into formula (19) yields an effective interparticle
spacing of $r_m \approx 0.48 pc$ for the fractal random stellar
medium. Let us compare it with traditional estimates of the mean
distance between stars. Published estimates of local stellar density
differ by about a factor of two. Thus the well-known handbook by
Allen and Cox (2004) gives $n \approx 0.141 pc^{-3}$. Bahcall and
Soneira (1986) normalize their Galaxy model to the local space
number density of $n(R_0) \approx 0.13 pc^{-3}$. Yanny and Gardner
(2013) estimate the local stellar number density to be $n \approx
0.109 ± 0.002 pc^{-3}$. When building their model of the Galaxy,
Roben et al. (2003) adopt the local mass density of $\rho \approx
0.076$ $M_0 pc^{-3}$, which, if converted into stellar number
density, also agrees with the above estimates. Substituting the
extreme values $0.109$ and $0.141 pc^{-3}$ into classical formula
(20) yields an average distance between the stars of $r_m \approx
1.16$ and $1.06 pc $, respectively, which is almost twice greater
than our preliminary estimate obtained with the allowance for the
fractal structure of the distribution of FG-type dwarfs.

We estimate the contribution of FG-type dwarfs to the total local
stellar density to be $0.03 - 0.04 pc^{-3},$ i.e., from 40 to 25\%.
If we assume that the fractal space distribution of FG-type dwarfs
is representative of that of all stars in the solar neighborhood
then to more accurately estimate the effect of fractality in the
full sample of disk stars the parameter h in formula (19) for the
effective distance between the stars should be increased
proportionally by a factor of 2.5 - 4 compared to our sampled value
of $h \approx 1.644$, i.e., up to $h \approx 4 - 7$. In this case
more realistic estimates of the effective interparticle spacing
$r_m$ for the entire local star sample decrease to 0.27 and 0.15 pc,
respectively, which is about a factor of four to six less than the
observed mutual distances between stars in the nearest solar
neighborhood. This result is due to the fact that fractal model
takes into account the gravitational clustering ("clumpiness") of
the stellar medium in regions with the radii of several tens of pc.

The characteristic time scale of the fluctuations of irregular
forces that are due to stellar encounters is on the order of $t_m
\approx r_m / v_m$, where $v_m$ is the mean relative velocity of
particles. Given that for a fractal medium $r_m$ can be almost by
one order of magnitude smaller than the corresponding value for a
uniform Poisson medium, it may be assumed that in fractal media
diffusion processes in the velocity space should play more important
part than classical diffusion and dynamic friction. This problem
requires special analysis.

\sectionb{5}{DISCUSSION}
Numerical simulations of the evolution of a medium with the initial
conditions in the form of a Poisson distribution of gravitating
point masses demonstrate rather fast granulation (formation of
clumped structure) with the sizes of both the granules (clumps) and
the corresponding voids spanning the entire range of scale lengths
from very short to extremely long. It is also known that the gravity
force acting on a test particle and produced by a particular
distribution of masses is determined both by the immediate
environment of the particle and large-scale properties of the entire
system. We already mentioned above that the classical Poisson model
of the space distribution of masses implies the Holtsmark
distribution for the magnitude of the random force acting on a test
particle.

Unlike the uniform medium model used by Chandrasekhar (1943), the
fractal gravitating medium of Carpenter-de Vaucouleurs-Mandelbrot
contains multiscale correlation structures. These structures are
characterized by stochastic self similarity and power-law
conditional density distribution of the form of equation (1) inside
them (Mandelbrot, 1977). This circumstance, in principle, makes it
extremely difficult to analytically derive the random force
distribution law in the general form. However, Vlad (1994) used a
specially developed functional-and-integral approach to analyze
non-uniform Poisson statistics, and applied his results to fractal
structures generated by gravitational field. He analytically derived
the distribution of random gravitational force without considering
all real space density fluctuations. When computing the field-star
density variations as a function of distance r from the test
particle, Vlad used (like we did in our case) the average densities
in spheres of radius r ignoring morphological peculiarities of
particular structures (deviations from the mean distribution) which
are usually present on any scale length. This approach is justified
when studying a fractal structure and it allows such structures to
be described in terms of correlation functions (Blumenfeld, Boll,
1993). In this approximation Vlad (1994) used n-point correlation
functions to perform complex analytical manipulations to generalize
the Holtsmark distribution and derive the following exact formula
for the probability density $W(F)$ of random force magnitude:
$$
W(F)dF =
\frac{DB}{2}(Gm)^{\frac{D}{2}}F^{-\frac{D+2}{2}}exp[-B(Gm)^{\frac{D}{2}}F^{-\frac{D}{2}}]
dF          \eqno(21)
$$
Here $D$ is the fractal dimension of the medium; $Â$, a constant
that characterizes the average mass in a unit sphere, and $m$, the
average mass of gravitating bodies.

Like in the case of our formula (14), uniform distribution
corresponds to a fractal dimension of $D = 3$. Thus for real
distributions of gravitating points function $W(F)$ decreases slower
with F than the corresponding function for uniform medium considered
by Chandrasekhar (1947).

Formulas (14) and (21) fully coincide up to notation. This means
that the distribution of random force in fractal gravitating media
with a power-law conditional density variation is fully determined
by the modified distribution of the nearest neighbor distance.

\textbf {Acknowledgments}

The study of the kinetics of stellar systems was supported by the
Russian Foundation for Basic Research (grant no. 14-02-00472), and
theoretical computations were supported by the Russian Science
Foundation (grant no. 14-22-00041).

\References

\refb Allen, C.W. Cox, A.N. 2004, {\it Allen's Astrophysical
Quantities}, New York: IAP Press: Springer

\refb Bahcall, J. Soneira, R. 1986, Ann. Rev. Astron. Astrophys.,
24, 577

\refb  Blumenfeld, R. Ball,R.C. 1993, Phys. Rev E. 47, 2298

\refb Bottaccio,M.   Montuori,M. Pietronero,L.   Miocchi P.and
Capuzzo Dolcetta,R. 2003, Mem. S.A.It. Suppl., 1, 120

\refb Vlad, M.O. 1994, Astrophys. Space Sci., 218, 159

\refb  de Vaucouleurs, G. 1970, Science, 167, Is. 3922, 1203

\refb de Vaucouleurs, G. et al., 1991, {\it Third Reference
Catalogue of Bright Galaxies. Volume I: Explanations and references.
Volume II: Data for galaxies between 0h and 12h. Volume III: Data
for galaxies between 12h and 24h} Springer

\refb  Wu, K.S. Lahav, O. Rees, M. 1999, Nature, 397, Is. 6716, 225

\refb  Gouliermis, D.A.Hony, S. Klessen, R.S. 2014, MNRAS, 439, 3775

\refb Gurevich, L.E. 1954, {\it Evolyutsiya zvesdnykh system
(Evolution of stellar systems). - In: Problems of cosmogono}. V.2,
Moscow (in Russian)

\refb  de Vega, H. Sanchez, N. Combes, F. Astrophys. J. 1998, 500, 8

\refb  Efremov, Yu.N. Elmegreen, B.G. 1998, MNRAS 299, 588

\refb  Carpenter, E.F. 1938, Astrophys. J., 88, 344

\refb Larson, R.B. 1981, MNRAS, 194, 809

\refb  Mandelbrot, B.B. 1977, {\it Fractals: Form, Chance and
Dimension}, W.H. Freedman\&Co., San Francisco.

\refb  Nordstr\"{o}m, B. Mayor,  M. Andersen 2004, J. J., et al. ,
A\&A, 418, 989

\refb Ogorodnikov, K.F. 1958, {\it Dynamic of the Stellar Systems}
Moskow (in Russian)

\refb Robin, A. Reyle, C. Derriere S. 2003, Picaud S., A\&A, 409,
523

\refb S\'{a}nchez, N. A?ez, N. Alfaro E.J.  and Crone Odekon, M.
 strophys. J. 2010, 720, 541

\refb Spitzer, L. Shvarzhild, M. 1953, Astrophys. J., 118, 1, 106

\refb Semelin, B. Combes, F. 2000, A\&A, 360,1096

\refb Semelin, B. Combes, F. 2002, A\&A, 387, 98

\refb Holmberg, J. Nordstr\"{o}m, B. Andersen, J. 2007, A\&A, 475,
519 )

\refb Holmberg, J. Nordstr\"{o}m, B.  Andersen, J. 2009, A\&A, 501,
941

\refb Chandrasekhar, S. 1943, Rev. Mod. Phys., 15, 1

\refb Chandrasekhar, S.  1943, {\it Stochastic Problems in Physics
and Astronomy}, Reviews of Modern Physics, 15, 1-89

\refb Chumak, O.V. Rastorguev, A.S. 2015, Baltic Astron., 24, 1, 302

\refb Chumak, O.V.  Chumak, Z.N. 1976, Tr. Astrofiz. Inst., Akad.
Nauk Kaz. SSR, 28, 81 (in Russian)

\refb Chumak, O.V.  Rastorguev, A.S. 2014, Pis'ma Astron. Zh., 40,
8, 517

\refb Elias, F.  Alfaro, E.J. Cabrero-Ca\~{n}o, J.  2009, MNRAS,
397, 2

\refb Elmegreen,D.B. Elmegreen,  B.G.Adamo A.  et al., 2014,
Astrophys. J. Letters, 787, L.15

\refb Yanny, B. Gardner,  S. 2013, Astrophys.J., 777, art. id.91

\end{document}